\documentclass[twocolumn,showpacs,superscriptaddress,aps,prl]{revtex4-1}
\usepackage{graphicx}
\usepackage{epstopdf}
\usepackage[usenames, dvipsnames]{xcolor}
\usepackage[caption=false,subrefformat=parens,labelformat=parens]{subfig}
\usepackage{hyperref}
\hypersetup{
    colorlinks=true,     
    linkcolor=blue,      
    citecolor=blue,      
    filecolor=blue,      
    urlcolor=blue        
}
\usepackage{amsmath}
\usepackage{pgffor}

\begin{document}

\newcommand{\fetese}{FeTe$_{0.55}$Se$_{0.45}$}
\newcommand{\feteten}{Fe$_{1.1}$Te}
\newcommand{\fetesix}{Fe$_{1.06}$Te}
\newcommand{\dg}{$^{\circ}$}
\newcommand{\appropto}{\mathrel{\vcenter{
  \offinterlineskip\halign{\hfil$##$\cr
    \propto\cr\noalign{\kern2pt}\sim\cr\noalign{\kern-2pt}}}}}

\newcommand{\bQ}{\mbox{\boldmath$Q$}}
\newcommand{\bq}{\mbox{\boldmath$q$}}
\newcommand{\br}{\mbox{\boldmath$r$}}
\newcommand{\btau}{\mbox{\boldmath$\tau$}}
\newcommand{\bstau}{\mbox{\boldmath$\scriptscriptstyle \tau$}}
\newcommand{\bxi}{\mbox{\boldmath$\xi$}}

\title{``Forbidden'' phonon: dynamical signature of bond symmetry breaking in the iron chalcogenides}

\author{David M. Fobes}
\email{dfobes@bnl.gov}
\affiliation{CMPMSD, 
 Brookhaven National Laboratory, Upton, NY 11973 USA}

\author{Igor A. Zaliznyak}
\email{zaliznyak@bnl.gov}
\affiliation{CMPMSD, 
 Brookhaven National Laboratory, Upton, NY 11973 USA}

\author{John M. Tranquada}
\affiliation{CMPMSD, 
 Brookhaven National Laboratory, Upton, NY 11973 USA}

\author{Zhijun Xu}
\affiliation{CMPMSD, 
 Brookhaven National Laboratory, Upton, NY 11973 USA}

\author{Genda Gu}
\affiliation{CMPMSD, 
 Brookhaven National Laboratory, Upton, NY 11973 USA}

\author{Xu-Gang He}
\affiliation{CMPMSD, 
 Brookhaven National Laboratory, Upton, NY 11973 USA}

\author{Wei Ku}
\affiliation{CMPMSD, 
 Brookhaven National Laboratory, Upton, NY 11973 USA}

\author{Yang Zhao}
\affiliation{NIST Center for Neutron Research, National Institute of Standards and Technology, Gaithersburg, Maryland 20899, USA}
\affiliation{DMSE, 
University of Maryland, College Park, MD 20742 USA}

\author{Masaaki Matsuda}
\affiliation{QCMD, 
 Oak Ridge National Laboratory, Oak Ridge, TN 37831 USA}

\author{V. Ovidiu Garlea}
\affiliation{QCMD, 
 Oak Ridge National Laboratory, Oak Ridge, TN 37831 USA}

\author{Barry Winn}
\affiliation{QCMD, 
 Oak Ridge National Laboratory, Oak Ridge, TN 37831 USA}

\date{\today}

\begin{abstract}
Investigation of the inelastic neutron scattering spectra in Fe$_{1+y}$Te$_{1-x}$Se$_{x}$ near a signature wave vector $\mathbf{Q} = (1,0,0)$ for the bond-order wave (BOW) formation of parent compound Fe$_{1+y}$Te \cite{FobesPRL2014} reveals an acoustic-phonon-like dispersion present in all structural phases. While a structural Bragg peak accompanies the mode in the low-temperature phase of Fe$_{1+y}$Te, it is absent in the high-temperature tetragonal phase, where Bragg scattering at this $\mathbf{Q}$ is forbidden by symmetry. Notably, this mode is also observed in superconducting \fetese{}, where structural and magnetic transitions are suppressed, and no BOW has been observed. The presence of this ``forbidden'' phonon indicates that the lattice symmetry is dynamically or locally broken by magneto-orbital BOW fluctuations, which are strongly coupled to lattice in these materials.
\end{abstract}


\maketitle

Since the discovery of high-temperature superconductivity (HTSC) in cuprates, elucidation of the connection between the electronic and lattice degrees of freedom has been of considerable interest in regard to the driving mechanisms behind HTSC. The iron-based superconductors (FeSCs) share many similarities with the cuprates; both have parent phases featuring antiferromagnetic (AFM) ordering, structural distortions, strong magnetic fluctuations and broken electronic symmetry \cite{FobesPRL2014,Scalapino2012,DaiHuDagotto2012,LumsdenChristianson2010,Fisher_RPP2011}. Understanding the complex lattice dynamics in FeSCs is of critical importance for understanding the connection between these different orders, their relation to the superconductivity, and the connection between the two types of HTSCs. The compounds in the iron-chalcogenide series, featuring the simplest structure and the strongest electronic correlation among the Fe-HTSC, provide a good opportunity to study these dynamics.

The iron-chalcogenides Fe$_{1+y}$Te$_{1+x}$Se$_{x}$, with a maximum $T_{c}$ of $\sim14.5$~K at optimal doping, consist of a continuous stacking of Fe square-lattice layers, separated by two half-filled chalcogen (Te,Se) layers \cite{Yeh2008,Hsu2009,Wen2011}. Initially predicted by band structure calculations to be a metal \cite{Singh2012}, the non-superconducting parent material Fe$_{1+y}$Te instead exhibits non-metallic character in resistivity, indicative of charge carrier incoherence near the Fermi level at high temperatures \cite{LiuPRB2009,Chen2009,HuPetrovic2009}. Large local magnetic moments of about 4~$\mu_{B}$, which indicate full involvement of three electronic bands, are revealed by Curie-Weiss behavior in magnetic susceptibility \cite{Chen2009,HuPetrovic2009,Zaliznyak_PRB2012}; nevertheless, angle-resolved photoemission (ARPES) studies show significant spectral weight near the Fermi energy \cite{Xia2009,ZhangFeng2010,Liu_Shen2013}.

These electronic and magnetic properties of Fe(Te,Se) are very sensitive to non-stoichiometric Fe at interstitial sites, particularly evident in the parent compound \cite{McQueen_PRB2009,Bao2009,Rodriguez2011,Li2009,Martinelli2010,Stock2011,Rossler2011,Liu2011}; at low concentrations, a first-order magnetostructural transition is observed from paramagnetic tetragonal ($P4/nmm$) to monoclinic ($P2_{1}/m$) with bicollinear AFM order (propagation vector $\mathbf{q} = (0.5,0,0.5)$) and metallic resistivity \cite{Bao2009,Li2009,Martinelli2010,Rodriguez2011}. For intermediate $0.06 \lesssim y \lesssim 0.12$ the magneto-structural transition splits into a sequence of transitions. Our recent neutron studies of Fe$_{1+y}$Te in the intermediate range uncovered evidence that the lowest-temperature transition coincides with the formation of a bond-order wave (BOW), indicative of ferro-orbital order in the ground state \cite{FobesPRL2014}, which stabilizes the bicollinear AFM order in the low-$T$ phase, common to $y \lesssim 0.12$.

In this letter, we report inelastic neutron scattering measurements on Fe$_{1+y}$Te$_{1-x}$Se$_{x}$ samples, aimed at elucidating the dynamics associated with the newly discovered BOW state. We have studied two compositions of the Fe$_{1+y}$Te parent compound, as well as the optimally-doped superconductor FeTe$_{1-x}$Se$_{x}$ ($x=0.45$). In all three samples, we observe an acoustic-phonon-like mode that appears to disperse out of the $\mathbf{Q} = (1,0,0)$ reciprocal lattice point, even in the tetragonal structure where such a Bragg peak is forbidden by crystal symmetry; only a weak reflection, manifest of the BOW state, develops at this wave vector in the parent compounds at low $T$. At all temperatures, the mode appears to be ungapped and sharp within instrumental resolution, demonstrating it is not a conventional soft mode; furthermore, we have confirmed that it is not a result of magnetic spin-flip scattering. The presence of this phonon suggests a dynamical breaking of crystal symmetry, potentially related to the magneto-orbital BOW fluctuations in these materials.

Neutron scattering measurements were carried out using the Hybrid Spectrometer (HYSPEC) \cite{Zaliznyak2005150,Winn_EJC2015} at the Spallation Neutron Source, Oak Ridge National Laboratory, and polarized neutron measurements were performed on the Double Focusing Triple-Axis Spectrometer (BT-7) \cite{Lynn_JResNIST2012} at the NIST Center for Neutron Research and the Polarized Triple-Axis Spectrometer (HB-1) at the High-Flux Isotope Reactor (HFIR) at Oak Ridge National Laboratory ($E_{f} = 14.7$~meV). The three following samples were investigated: \fetesix{}, \feteten{}, and \fetese{}. \fetesix{} consists of two co-aligned single crystals ($m_{\mathrm{total}}\approx24$~g) with a mosaic of 2.7\dg{} full width at half maximum (FWHM). \feteten{} is a single crystal ($m = 18.45$~g) with a mosaic of 2.2\dg{} FWHM. \fetese{} is a single crystal ($m=23.4$~g) with a mosaic of 2.2\dg{} FWHM. All samples were grown by the horizontal Bridgman method \cite{Wen2011}, and mounted on an aluminum holder. Measurements of scattering in the $(h,k,0)$ and the $(h,0,l)$ plane were obtained by aligning crystals with $c$-axis, or $b$-axis vertical, respectively

\begin{figure}[tbp]
\vspace{-0.1in}
\subfloat{\includegraphics[width=0.99\linewidth]{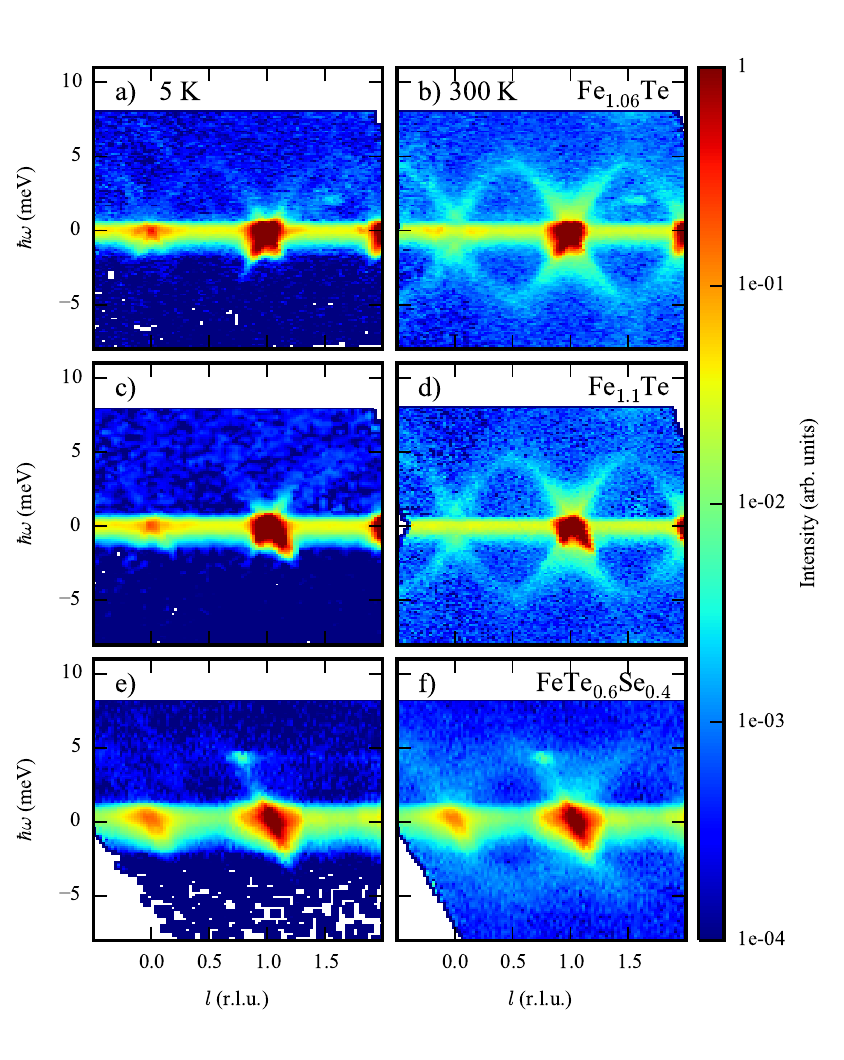}
\label{h0lvE:a}}
\subfloat{\label{h0lvE:b}}\subfloat{\label{h0lvE:c}}\subfloat{\label{h0lvE:d}}\subfloat{\label{h0lvE:e}}\subfloat{\label{h0lvE:f}}
\vspace{-0.3in}
\caption{(Color Online) Neutron scattering intensity along $(1,0,l)$ as a function of energy transfer $\hbar\omega$, showing the phonon dispersions from (100), (101) and (102) at 5~K (left column) and 300~K (right column) for (a--b) \fetesix{}, (c--d) \feteten{}, and (e--f) \fetese{}.}
\vspace{-0.2in}
\label{h0lvE}
\end{figure}

In Fig.~\ref{h0lvE}, we present inelastic neutron scattering intensity, which reveals acoustic-phonon-like dispersions along $(1,0,l)$, as a function of energy transfer $\hbar\omega$, for all three samples. At 5~K the structure in the Fe$_{1+y}$Te samples is monoclinic ($P2_{1}/m$) \cite{FobesPRL2014}, and acoustic phonons dispersing out of the (100) and (101) Bragg peaks present at these positions would not be unexpected (Fig.~\ref{h0lvE:a},~\ref{h0lvE:c}). However, at 300~K the structure is tetragonal (P4/$nmm$) \cite{Bao2009} and the (100) Bragg reflection is symmetry-forbidden; nevertheless, a gapless, acoustic-phonon-like mode is still observed dispersing out of the forbidden Bragg position (Fig.~\ref{h0lvE:b},~\ref{h0lvE:d}). 
A similar mode is also observed in superconducting \fetese{} throughout the whole temperature range (Figs.~\ref{h0lvE:e}--\ref{h0lvE:f}). In \fetese{} the structural and magnetic transitions observed in the parent compound are suppressed, and it is therefore tetragonal at all temperatures, so that Bragg scattering at (100) is never allowed. In Figs.~\ref{h0lvE:e}--\ref{h0lvE:f}, residual elastic scattering appears to be present at (100), but elastic slices (data not shown) reveal an unusual structure to the (100) peak, which changes depending on the incident energy, suggesting that these elastic features are a result of multiple scattering, not uncommon in samples of this size. The observed forbidden mode appears to be ungapped within the experimental limit, $\lesssim 1$~meV, mainly imposed by the presence of this spurious double scattering.

\begin{figure}[tbp]
\subfloat{
\includegraphics[width=0.99\linewidth]{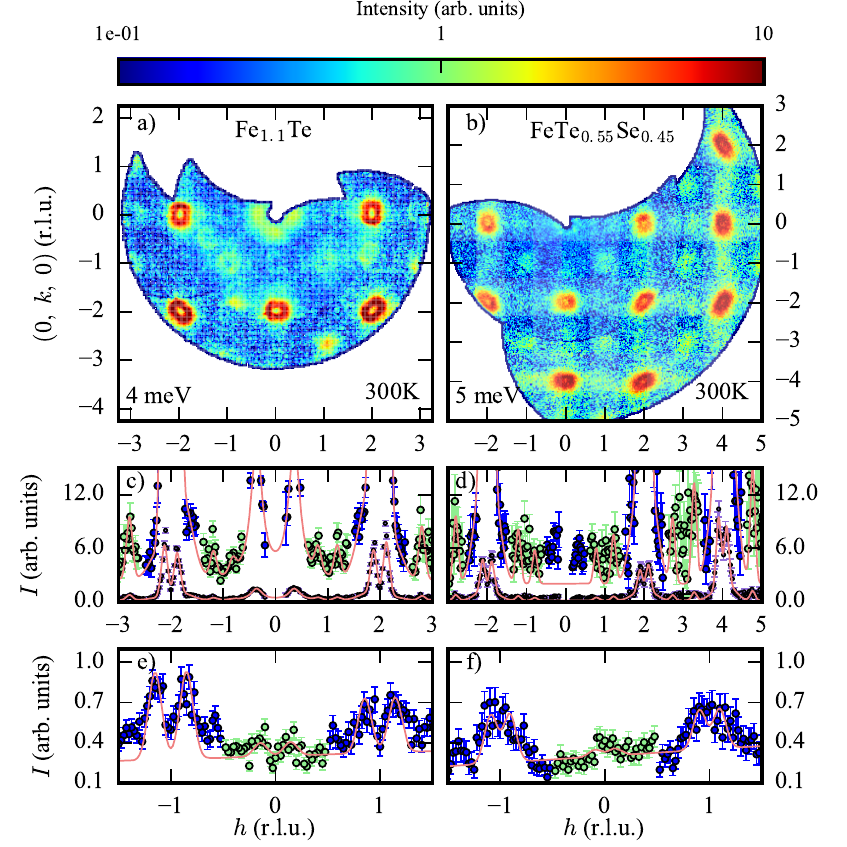}
\label{samples:a}}
\subfloat{\label{samples:b}}\subfloat{\label{samples:c}}\subfloat{\label{samples:d}}\subfloat{\label{samples:e}}\subfloat{\label{samples:f}}\subfloat{\label{samples:g}}\subfloat{\label{samples:h}}
\vspace{-0.0in}
\caption{(Color Online) The inelastic neutron scatering intensity maps at $\hbar\omega=4.0(5)$~meV in \feteten{} (a) and $\hbar\omega=5(1)$~meV in \fetese{} (b) in the $(h,k,0)$ zone, measured on HYSPEC using $E_{i} = 24$~meV and $E_{i} = 50$~meV, respectively. Panels (c) and (d) show the longitudinal and (e) and (f) the transverse line cuts of the data in (a) and (b), respectively, across the symmetry equivalent forbidden Bragg positions. The lines are Gaussian fits, revealing the longitudinal phonon velocities near $(2,0,0)$, $v_L = 36(2)$~meV/r.l.u. in \feteten{} and $v_L = 44(8)$~meV/r.l.u. in \fetese{}, compared to $v = 21(1)$~meV/r.l.u. and $v = 22(2)$~meV/r.l.u., respectively, for the forbidden mode [r.l.u. is in units of $a^* =1.645(5) \AA^{-1}$].}
\label{samples}
\vspace{-0.2in}
\end{figure}

In Fig.~\ref{samples} we show constant-energy inelastic data covering a large region of the $(h,k,0)$ plane for the \feteten{} and \fetese{} samples. The data reveal the presence of ring-like contours of inelastic scattering intensity consistent with the dispersion of excitations around the forbidden Bragg peaks at the symmetry equivalent positions in different Brillouin zones, $(\pm 2n\pm 1,0,0)$, $(0,-2n-1,0)$, $n=0,1,2$. The line cuts presented in Fig.~\ref{samples:c}, \ref{samples:d}, reveal both the longitudinal acoustic phonon branch near $(\pm 2n,0,0)$ and the small ``forbidden mode'' peaks near $(\pm 2n + 1,0,0)$, whose intensity distribution with respect to the wave vector direction is consistent with that of a longitudinal acoustic phonon, as discussed in more detail below. However, the forbidden peak position reveals a dispersion with velocity, $v = 21(1)$~meV/r.l.u. in \feteten{} and $v = 22(2)$~meV/r.l.u. in \fetese{}, which is markedly (nearly twice) slower than the respective longitudinal phonon velocities, $v_L \approx 36(2)$~meV/r.l.u. and $v_L = 44(8)$~meV/r.l.u., and is close to that of the transverse acoustic modes. Hence, the combination of polarization and dispersion of the forbidden mode is inconsistent with that expected for a phonon in an ideal lattice. In addition, the forbidden phonon intensities at symmetry equivalent positions do not display the expected scaling with wave vector, $\sim Q^2$; one should expect that the position-normalized phonon intensity at symmetry-equivalent positions, $I/Q^2$, should be constant as a function of $Q$, whereas we observe a significant decrease \cite{supplement}.

\begin{figure}[tbp]
\subfloat{
\includegraphics[width=0.99\linewidth]{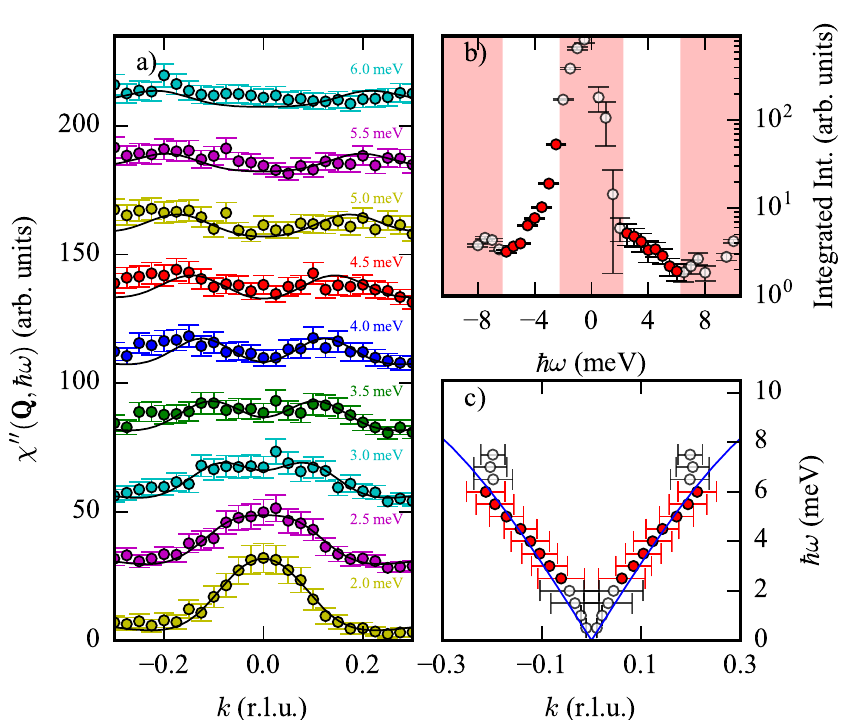}
\label{analysis:a}}
\subfloat{\label{analysis:b}}\subfloat{\label{analysis:c}}
\vspace{-0.1in}
\caption{(Color Online) (a) $\chi^{\prime\prime}(\mathbf{Q}, \hbar\omega)$ line scans along $(1,k,0)$ at energy transfers between 2~meV and 6~meV at 300~K of the \fetese{} sample, fitted with two gaussians symmetric around $k=0$. Data are shifted for clarity. (b) Integrated peak intensity as a function of energy, obtained from fitting. (c) Phonon dispersion in energy fitted to $\hbar\omega = v\left|\sin(\pi k/2)\right|$ (solid line). Shaded regions in (b) and open symbols in (b) and (c) indicate regions where fitting is least reliable.}
\label{analysis}
\vspace{-0.2in}
\end{figure}

In Fig.~\ref{analysis}, we present a set of line-cuts along $(1,k,0)$ of $\chi^{\prime\prime}(\mathbf{Q}, \hbar\omega)$ in \fetese{} at multiple energy transfers, 2~meV $\leq \hbar \omega \leq$ 6~meV (Fig.~\ref{analysis:a}), which we use for an analysis of the transverse acoustic-phonon-like dispersion near (100) (Fig.~\ref{h0lvE:f}). Line cuts were fit to a two-gaussian function, where the gaussians were constrained to be symmetric around $k=0$. Figures~\ref{analysis:b}--\ref{analysis:c} show the results of fitting, the total integrated intensity and dispersion, respectively. The dispersion is fit to $\hbar\omega = v \left|\sin(\pi k)\right|$, yielding the acoustic velocity $\pi v = 32(2)$~meV/r.l.u.

The neutron scattering cross-section of a phonon mode with index $\nu$ and wave vector ${\bf q}$ measured at a wave vector ${\bf Q} = {\bf q} + \btau$ near the reciprocal lattice vector $\btau$ and at a temperature $T$,
\begin{align}
\label{scatt-eq} \frac{d^{2}\sigma}{dE d\Omega} = & N \frac{k_f}{k_i} \left|{\bf Q} \cdot \mathbf{g}^{\nu}_{\bf Q}\right|^2 \frac{\hbar}{2 \omega_\nu({\bf q})} \frac{ \delta\left(\hbar\omega - \hbar\omega_\nu({\bf q}) \right)}{1-e^{-\hbar\omega/T}},
\end{align}
is proportional to the square of the structure factor,
\begin{equation}
\label{structfac} \mathbf{g}^{\nu}_{\bf Q}  = \sum_{j} \frac{{b}_j}{\sqrt{M_j}} e^{-W_j({\bf Q})} e^{i{\bf Q} \cdot {\bf r}_j} \bxi^\nu_{j}({\bf q}) , 
\end{equation}
where $e^{-2W_j({\bf Q})}$ and $b_j$ are the Debye-Waller factor and the scattering length of an atom of mass $M_j$ at a position ${\bf r}_j$ in the unit cell ($N$ is the number of unit cells);  $\omega_\nu({\bf q}) = \omega_\nu({\bf q}+\btau)$ and $\bxi^\nu_{j}({\bf q}) = \bxi^\nu_{j}({\bf q} + \btau)$ are the mode frequency and polarization vectors, which are given by the eigenvalues and the eigenvectors of the dynamical matrix \cite{BrokhouseIyengar_PhysRev1959,ElliottThorpe}. Due to lattice periodicity, ${\bf q}$ can be constrained to the first Brillouin zone.

In the long-wavelength limit of acoustic phonons, all atoms in the unit cell move uniformly together, and the magnitude of the phonon structure factor approaches that of the static structure factor at the Bragg position, $\btau$, from which the dispersion originates,
$|\mathbf{g}^{\nu}_{\bf Q}\cdot{\bf Q}|^2 \stackrel{\longrightarrow}{\scriptscriptstyle {\bf Q}\rightarrow\bstau} (\tau^2/M) |F(\btau)|^2 \cos^2\beta$, where $M$ is the sum of the atomic masses and $\beta$ is the angle that the phonon polarization makes with \btau. Therefore, in the situation where $\btau$ is a forbidden Bragg reflection, $|F(\btau)| = 0$ and acoustic phonon scattering is forbidden in this approximation. Using a Taylor expansion about ${\bf Q} = \btau +{\bf q}$ near $\btau$, we obtain, $|\mathbf{g}^{\nu}_{\bf Q}|^2 \lesssim \alpha q^2$. Thus, taking account of the linear dispersion of an acoustic mode at small $q$, the one-phonon scattering intensity converted to $\chi^{\prime\prime}({\bf Q}, \omega)$ by adjusting for the thermal balance factor in Eq. (\ref{scatt-eq}), which is proportional to $|\mathbf{g}^{\nu}_{\bf Q}|^2/\omega_\nu(q) \lesssim \alpha q /v$, should be decreasing to zero, at most linearly in $q$, as $q \rightarrow 0$ and $\hbar\omega \rightarrow 0$. This is in contradiction to what is seen in Fig.~\ref{analysis:b}, where the integrated intensity of $\chi^{\prime\prime}({\bf Q}, \omega)$, \emph{increases} as $\hbar \omega$ decreases. Analysis of \feteten\ and \fetesix\ data (Figs.~\ref{h0lvE:a}--\ref{h0lvE:d}) leads to similar conclusions \cite{supplement}.

\begin{figure}[tbp]
\vspace{0.1in}
\subfloat{
\includegraphics[width=0.99\linewidth]{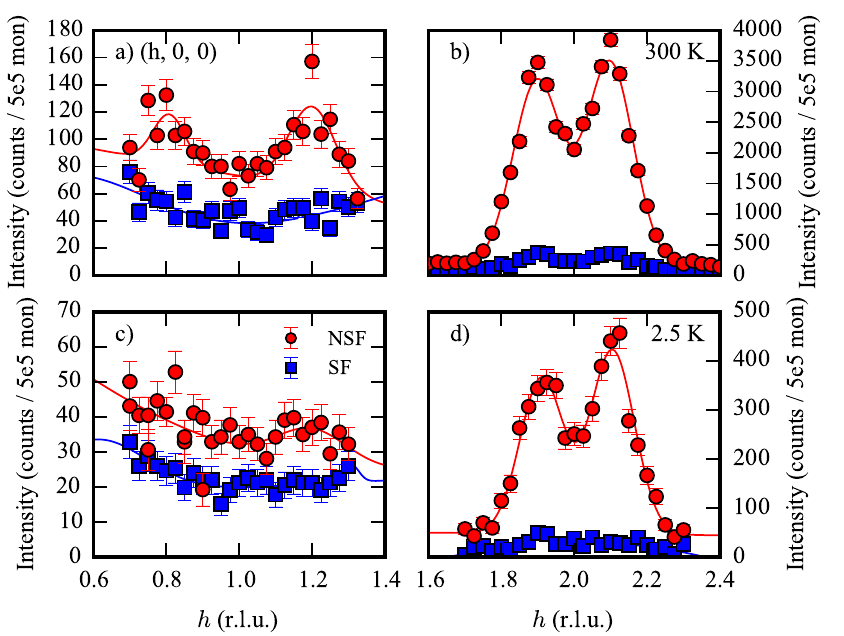}
\label{polarized:a}}
\subfloat{\label{polarized:b}}\subfloat{\label{polarized:c}}\subfloat{\label{polarized:d}}
\vspace{-0.1in}
\caption{(Color Online) Polarized inelastic neutron scattering line scans along (h,0,0) around (1,0,0) (left column) and (2,0,0) (right column) at 300~K (top row) and 2.5~K (bottom row) in the non-spin-flip (circle) and spin-flip (square) channels of the \fetesix{} sample, measured at BT-7.}
\label{polarized}
\vspace{-0.2in}
\end{figure}

In order to understand the possible origin of the observed ``forbidden phonon'' mode, we performed LDA frozen phonon calculations, which reveal a coupling between atomic displacements and a spin imbalance of neighboring Fe atoms, suggesting the possibility that a forbidden phonon mode could result from magnetic scattering induced by thermal atomic vibrations \cite{supplement}. To test this, polarized neutron measurements were performed on the \fetesix{} sample, as shown in Fig.~\ref{polarized}, where spin-flip (SF) and non-spin flip (NSF) scattering was measured at 300~K and 2.5~K using $^{3}$He polarizers and ${\bf Q}\parallel \mathbf{B}$ (guide field). With a median flipping ratio of $\sim30$ during these experiments, our results indicate a lack of SF magnetic scattering from either the expected (200) phonon or the ``forbidden'' (100) mode. A similar polarized neutron experiment was performed on the \fetese{} sample \cite{supplement} at HB-1, yielding consistent results. While the absence of spin-flip scattering indicates the mode primarily originates from atomic displacements, this does not exclude a scenario in which the atomic displacements originate from magnetic/orbital fluctuations, as we discuss below.

A previous example of a ``forbidden'' phonon was found in Fe$_{65}$Ni$_{35}$ invar, in which a transverse acoustic (TA) mode is observed in a position where it is forbidden by the scattering geometry, \textit{i.e.} ${\bf Q} \cdot \bxi({\bf q}) = 0$, cf. Eqs. (\ref{scatt-eq}--\ref{structfac}) \cite{Brown89}. An early explanation suggested this mode could result from the breaking of the cubic crystal symmetry of the dynamical matrix by slow local orthorhombic distortions \cite{Lipinski94}. However, this mode has several properties divergent from those expected for a conventional phonon. The mode does not exhibit the expected $Q^2$-dependence, but instead shows a decrease in intensity at higher $Q$. The mode also shows a significant \emph{decrease} in intensity at temperatures well above the magnetic ordering temperature ($T_{c} \approx 550$~K). Finally, there is a polarization ratio associated with this mode, where some contribution to the mode is structural and some is magnetic \cite{Brown96}. These results suggested that the scattering intensity is in part a result of magnetic scattering, but the strong coupling of the magnetic and lattice degrees of freedom also results in magnetically-driven structural phonon scattering.

The magnetic fluctuations in the iron chalcogenides are well known to be strong, even in the absence of long-range magnetic order in these materials \cite{Zaliznyak_PRL2011,Zaliznyak_PNAS2015}, and previous studies have emphasized the strong coupling between the electronic spin, orbital, and lattice degrees of freedom \cite{FobesPRL2014}. The location of the ``forbidden'' phonon, which we observe near the wave vector of ferro-orbital ordering in the parent FeTe material \cite{FobesPRL2014} clearly suggests involvement of the orbital degrees of freedom. This is consistent with the magneto-vibrational scenario set forth by our LDA frozen phonon calculation, similar to the scenario offered to explain the ``forbidden" phonon in Fe-Ni invar, but with an additional factor of the orbital hybridization, a crucial ingredient. Specifically, fluctuations of the orbital/magnetic nature associated with the spin imbalance between neighboring Fe atoms might lead to vibrations of forbidden character \cite{supplement}.

Recently, another example of a ``forbidden'' phonon has been observed in La$_{2-x}$Ba$_{x}$CuO$_{4}$ ($x=0.125$), where the mode has been attributed to CuO$_6$ octahedral tilt fluctuations \cite{PhysRevB.91.054521}. A possible electronic coupling for the observed mode is not ruled out. It could be that an acoustic phonon-like mode near a structurally forbidden Bragg reflection is a universal feature indicative of coupling between electronic and lattice degrees of freedom in both cuprates and iron chalcogenides.

In summary, we have observed an acoustic phonon-like mode dispersing from a position where Bragg scattering is forbidden by crystal symmetry in both the non-superconducting end-member and optimally-doped superconducting member of the iron-chalcogenide family. The mode intensity does not follow the expected behavior for phonon scattering near a forbidden Bragg reflection. Frozen phonon LDA calculations suggest that this mode might originate from slow electronic magnetic/orbital fluctuations associated with Fe spin moments leading to a dynamical breaking of the crystal unit cell symmetry.

\begin{acknowledgments}
Work at BNL was supported by Office of Basic Energy Sciences (BES), Division of Materials Sciences and Engineering, U.S. Department of Energy (DOE), under Contract No. DE-SC00112704. Research conducted at ORNL's Spallation Neutron Source and High Flux Isotope Reactor was sponsored by the Scientific User Facilities Division, Office of Basic Energy Sciences, US Department of Energy. We acknowledge the support of NIST, US Department of Commerce, in providing the neutron research facilities used in this work.
\end{acknowledgments}

\bibliographystyle{apsrev4-1}

\pagebreak
\widetext
\setcounter{page}{1}
\onecolumngrid

\begin{figure}[h] 
\vspace{-2cm}
\hspace*{-2cm}\includegraphics[scale=1.0]{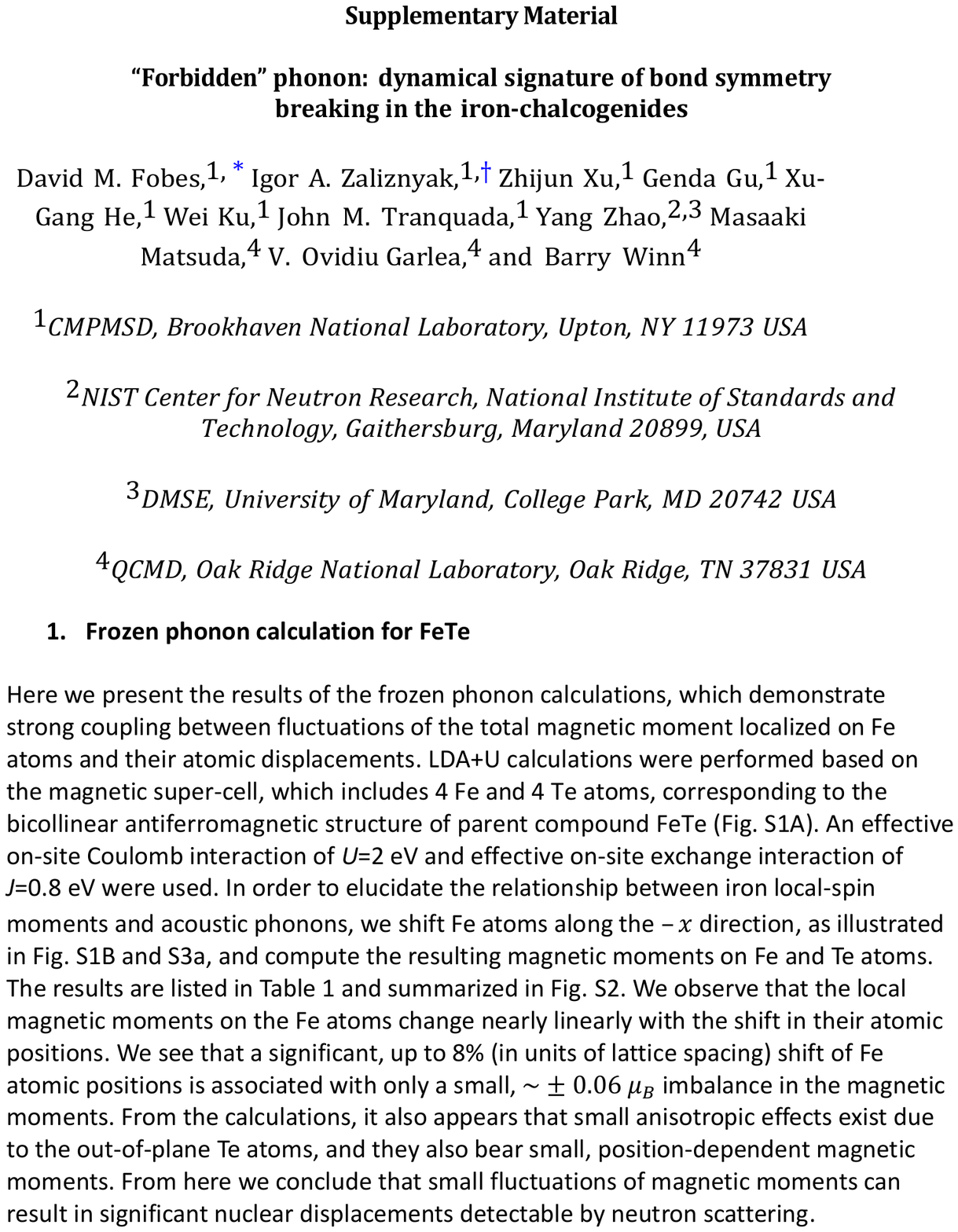}
\end{figure}

\newpage
\begin{figure}[h] 
\vspace{-2cm}
\hspace*{-2cm}\includegraphics[scale=1.0]{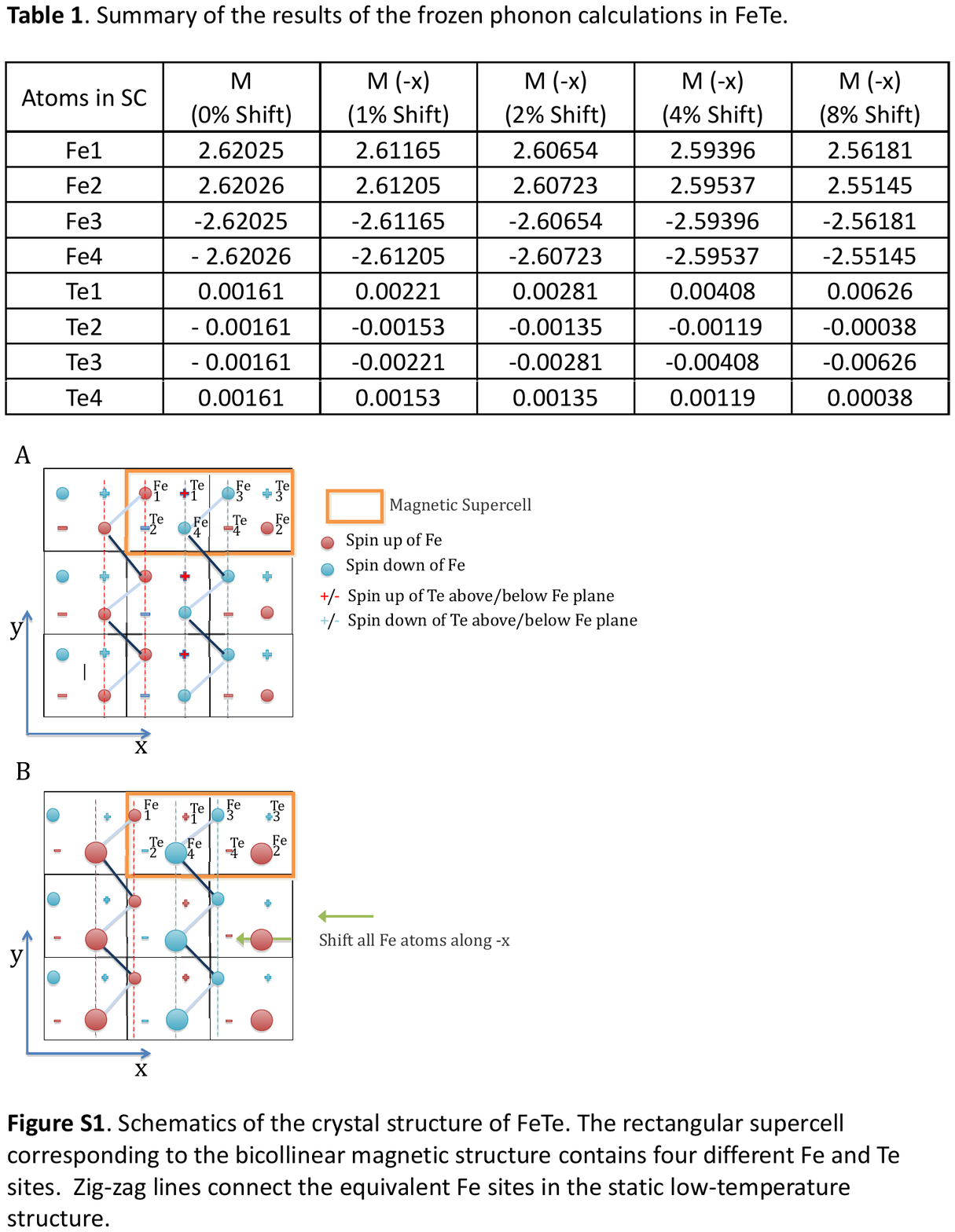}
\end{figure}

\newpage
\begin{figure}[h] 
\vspace{-2cm}
\hspace*{-2cm}\includegraphics[scale=1.0]{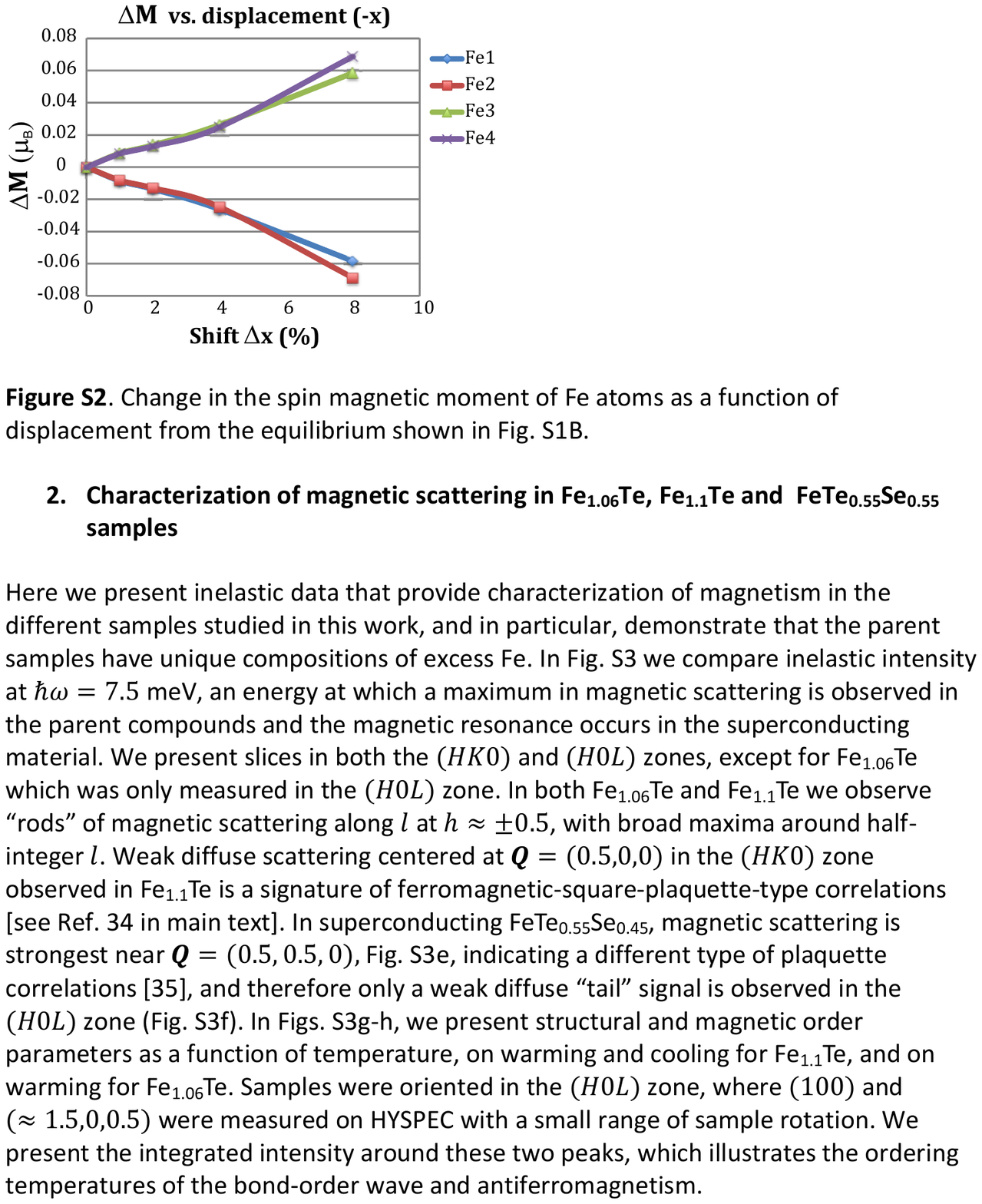}
\end{figure}

\newpage
\begin{figure}[h] 
\vspace{-2cm}
\hspace*{-2cm}\includegraphics[scale=1.0]{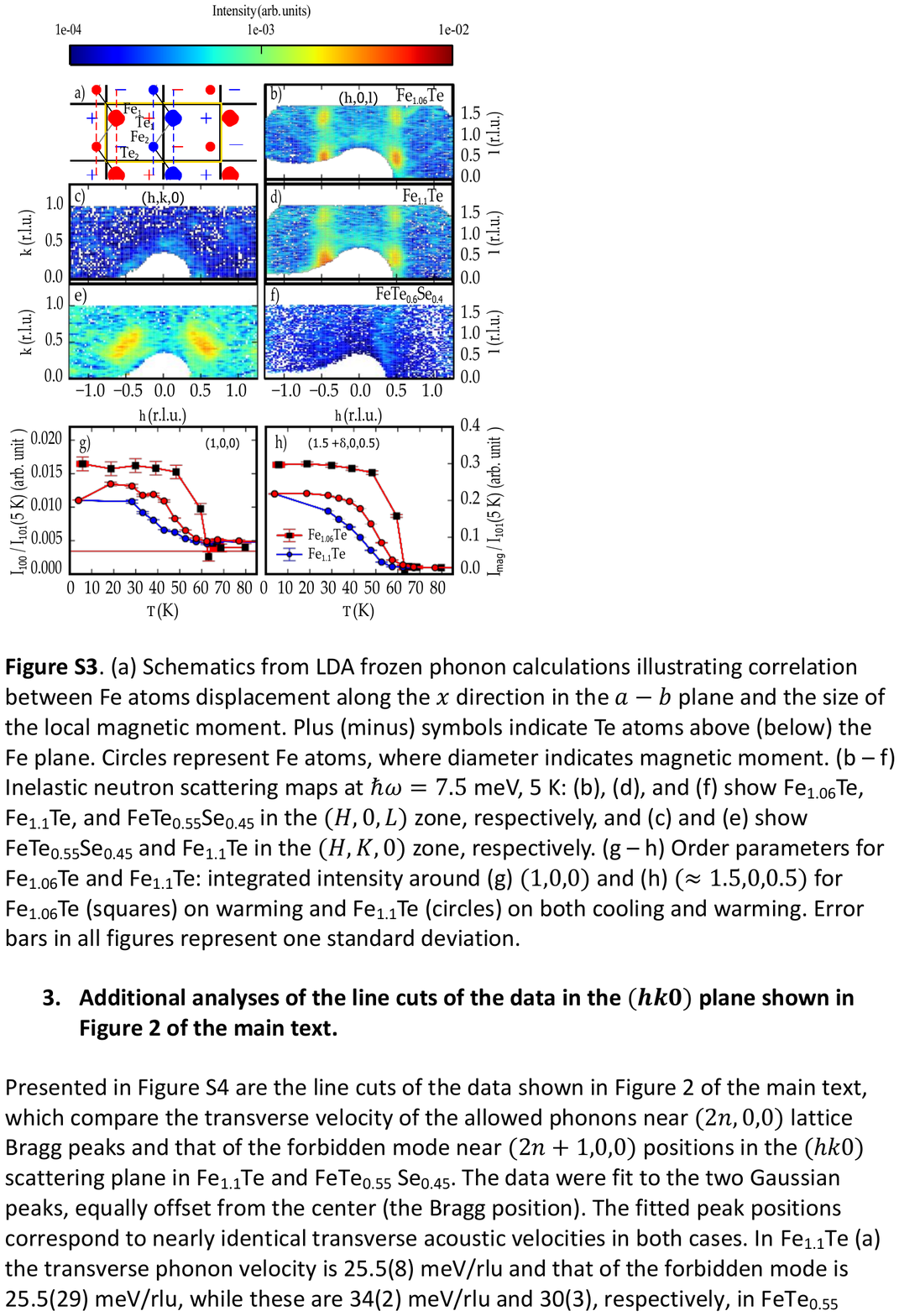}
\end{figure}

\newpage
\begin{figure}[h] 
\vspace{-2cm}
\hspace*{-2cm}\includegraphics[scale=1.0]{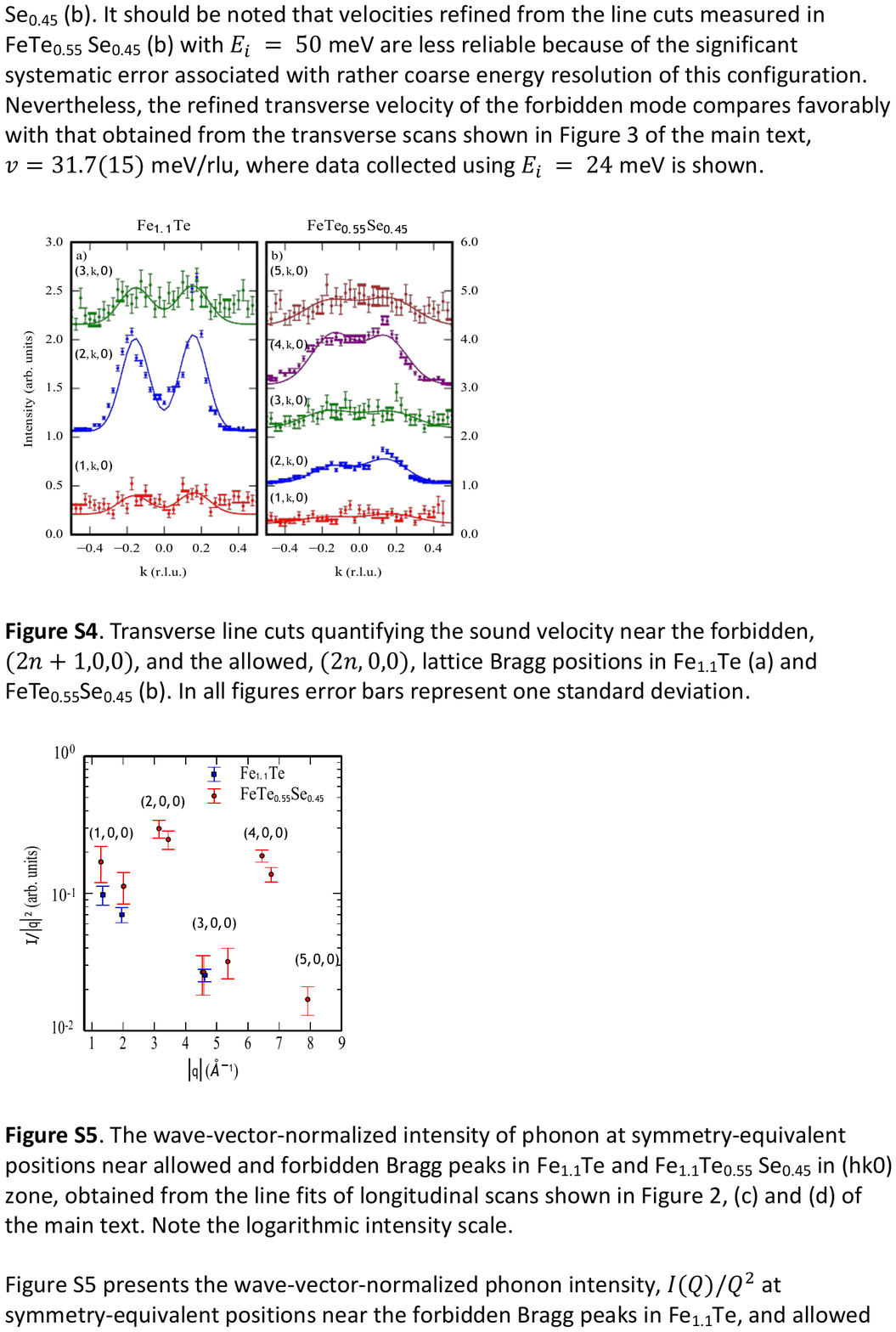}
\end{figure}

\newpage
\begin{figure}[h] 
\vspace{-2cm}
\hspace*{-2cm}\includegraphics[scale=1.0]{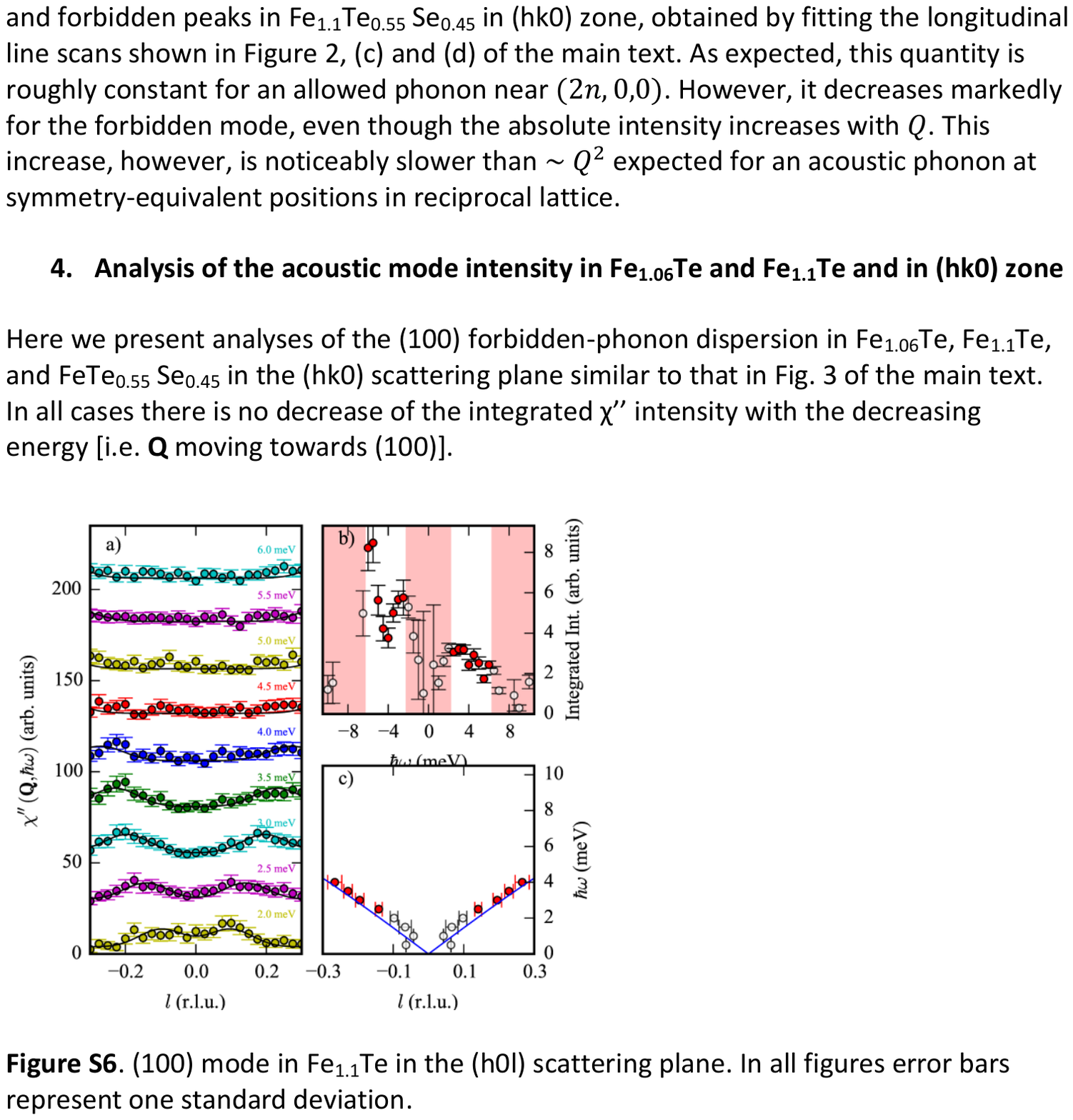}
\end{figure}

\newpage
\begin{figure}[h] 
\vspace{-2cm}
\hspace*{-2cm}\includegraphics[scale=1.0]{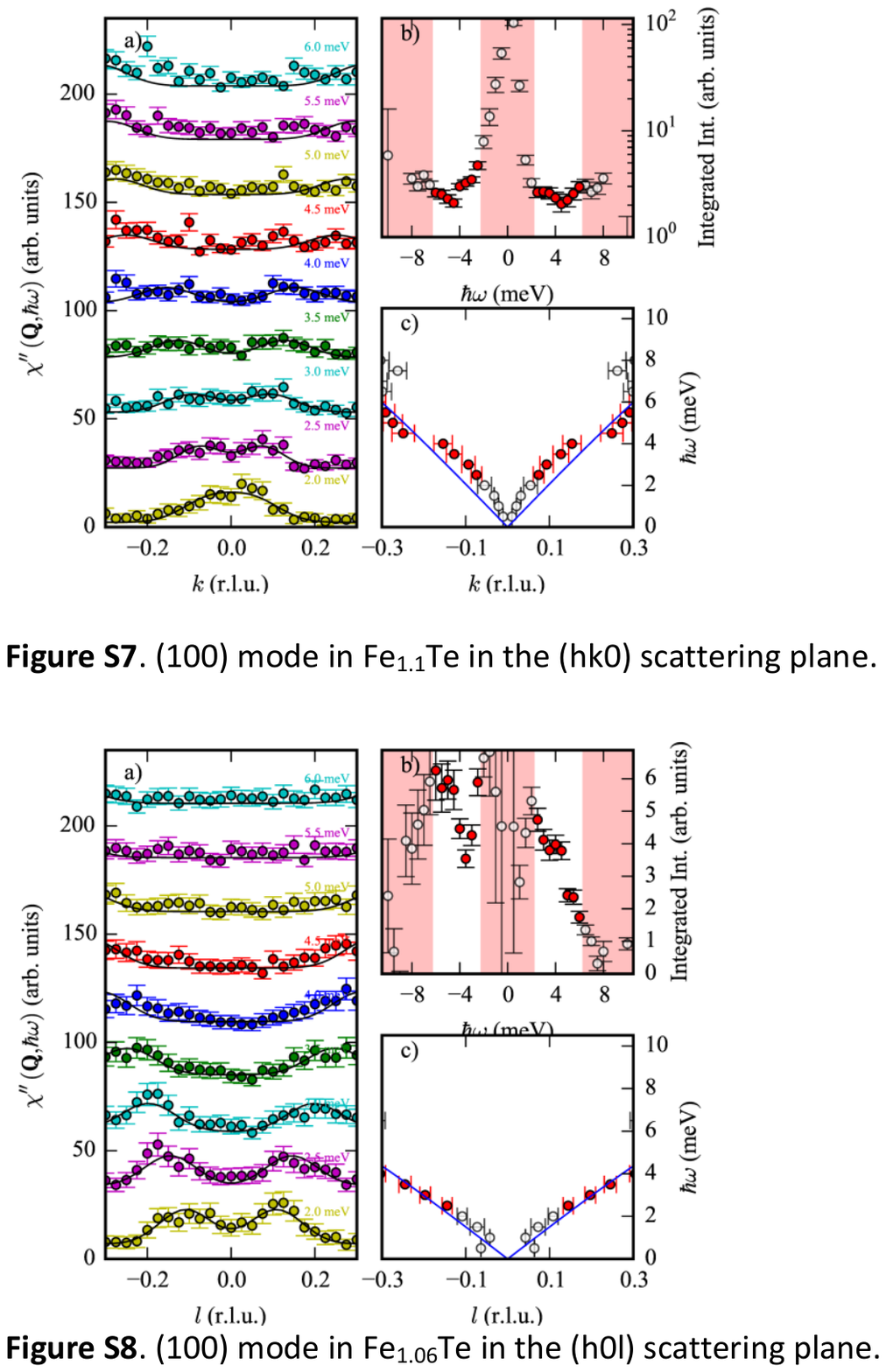}
\end{figure}

\newpage
\begin{figure}[h] 
\vspace{-2cm}
\hspace*{-2cm}\includegraphics[scale=1.0]{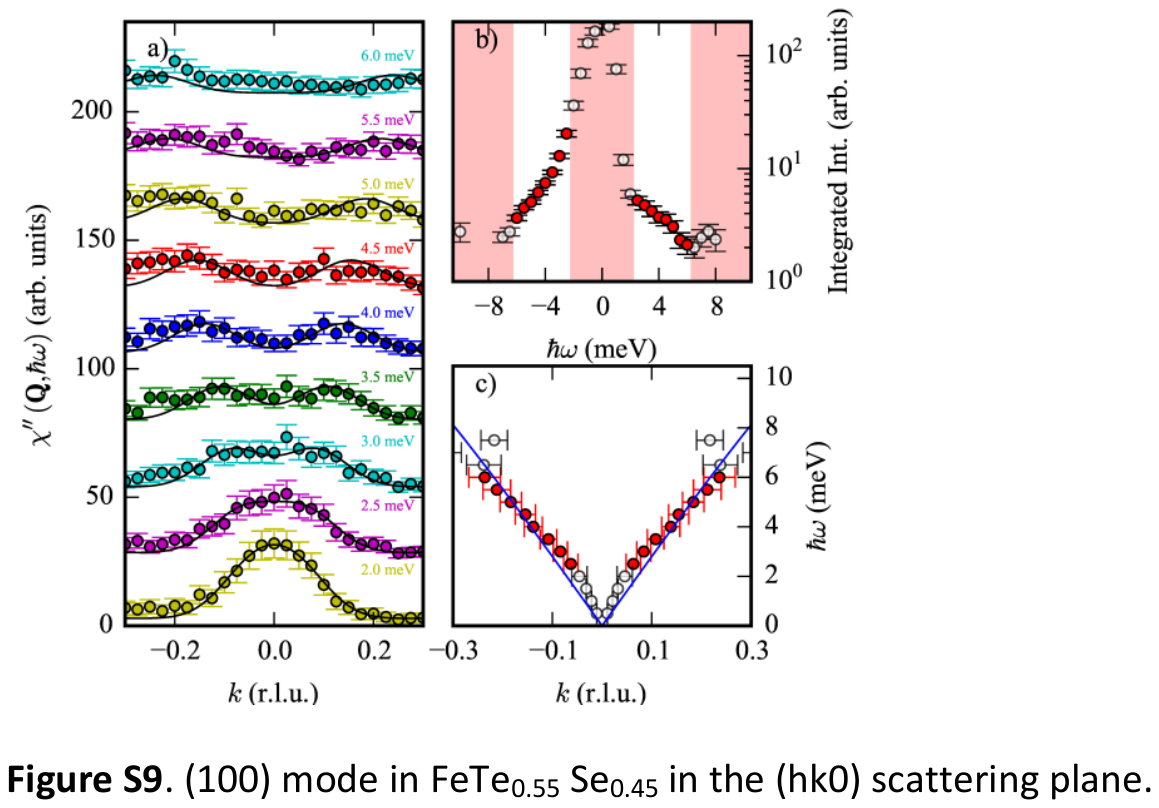}
\end{figure}

\end{document}